Article

# Exploring the Spatial Features of Electronic Transitions in Molecular and Biomolecular Systems by Swift Electrons

Ciro A. Guido,* Enzo Rotunno,* Matteo Zanfrognini, Stefano Corni, and Vincenzo Grillo

Cite This: https://dx.doi.org/10.1021/acs.jctc.1c00045

Read Online

ACCESS | Metrics & More | Article Recommendations | Supporting Information

**ABSTRACT:** We devise a new kind of experiment that extends the technology of electron energy loss spectroscopy to probe (supra-)molecular systems: by using an electron beam in a configuration that avoids molecular damage and a very recently introduced electron optics setup for the analysis of the outcoming electrons, one can obtain information on the spatial features of the investigated excitations. Physical insight into the proposed experiment is provided by means of a simple but rigorous model to obtain the transition rate and selection rule. Numerical simulations of DNA G-quadruplexes and other biomolecular systems, based on time dependent density functional theory calculations, point out that the conceived new technique can probe the multipolar components and even the chirality of molecular transitions, superseding the usual optical spectroscopies for those cases that are problematic, such as dipole-forbidden transitions, at a very high spatial resolution.

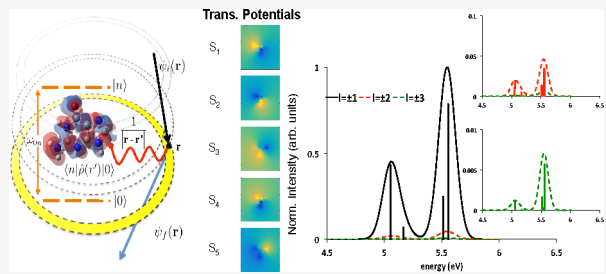

## ■ INTRODUCTION

Understanding the electronic structure of matter is a formidable task that largely made use of optical spectroscopies and their corresponding selection rules; indeed, when probing optical excitations with nanometer resolution, one can obtain information on their dynamics and interactions down to the atomic scale.[1,2]

The information acquired can range from the electronic structure and properties of a single molecule to the energy and electron transfer mechanism in complex systems, just to cite a few.[1,3] The origin of spectral lines is due to the absorption, emission, and scattering of a photon that modify the energy of the system, whereas the line shape can carry information about the dissipation of the energy absorbed, the interaction with the surroundings, and its influence in modulating the microscopic dynamics of chromophores.[1,4−6] However, not all of the electronic transitions can be probed in optical spectroscopic experiments because of different selection rules: being optically forbidden, the possibility to investigate the role of a given transition in the photophysical and/or photochemical activity of a molecular system is precluded. For instance, a long debate in the literature is still ongoing on the possible role of charge transfer (CT) states in photosynthetic mechanisms: being dark, it can only be indirectly probed.[7,8] On the other hand, electron-beam spectroscopes are now emerging as probing techniques to study optical excitations with combined space, energy, and time resolution:[9] nanophotonic structures and their detailed optical responses are now starting to be explored.[10] Between the different types of probe experiments that can be performed in trasmission electron microscopes (TEM) and scanning TEM (STEM), electron energy-loss spectroscopy (EELS) can provide insight into the properties of materials on the nanoscale:[11,12] it is widely used to identify chemical species with atomic resolution[13−15] through their characteristic high-energy core losses. Additionally, low-loss EELS can probe the spatial and spectral distributions of plasmons in metallic nanostructures,[16−22] and more recently, also phonons in polaritonic materials have been investigated.[23] The main advantage is the possibility to spatially map the fields associated with both bright and dark plasmonic resonances of a given nanostructure. Usually, EELS experiments produce swift electrons (from 30 to 300 keV typically) that interact with the sample exchanging energy and momentum. The loss function (i.e., the probability, per unit of transferred frequency, that the swift electron loses energy) is evaluated at the excitation energy of a given plasmonic resonance.[11] If spectroscopy carried out in the electron microscopes could be extended to the molecular and supramolecular systems, then this technique could be used not only to determine the overall morphology but also to follow the dynamics of electronic processes inside complex molecular aggregates: for instance, one could find at high spatial resolution where the different chromophores are located within the overall structure in proteins and pigment−protein complexes and then study the processes leading to energy and electron transfer. In this direction, very recent studies have shown applications of EELS to study vibrations in guanine crystals, resolving their characteristic C−H, N−H, and C═O vibrational signatures with no observable radiation damage.[24]

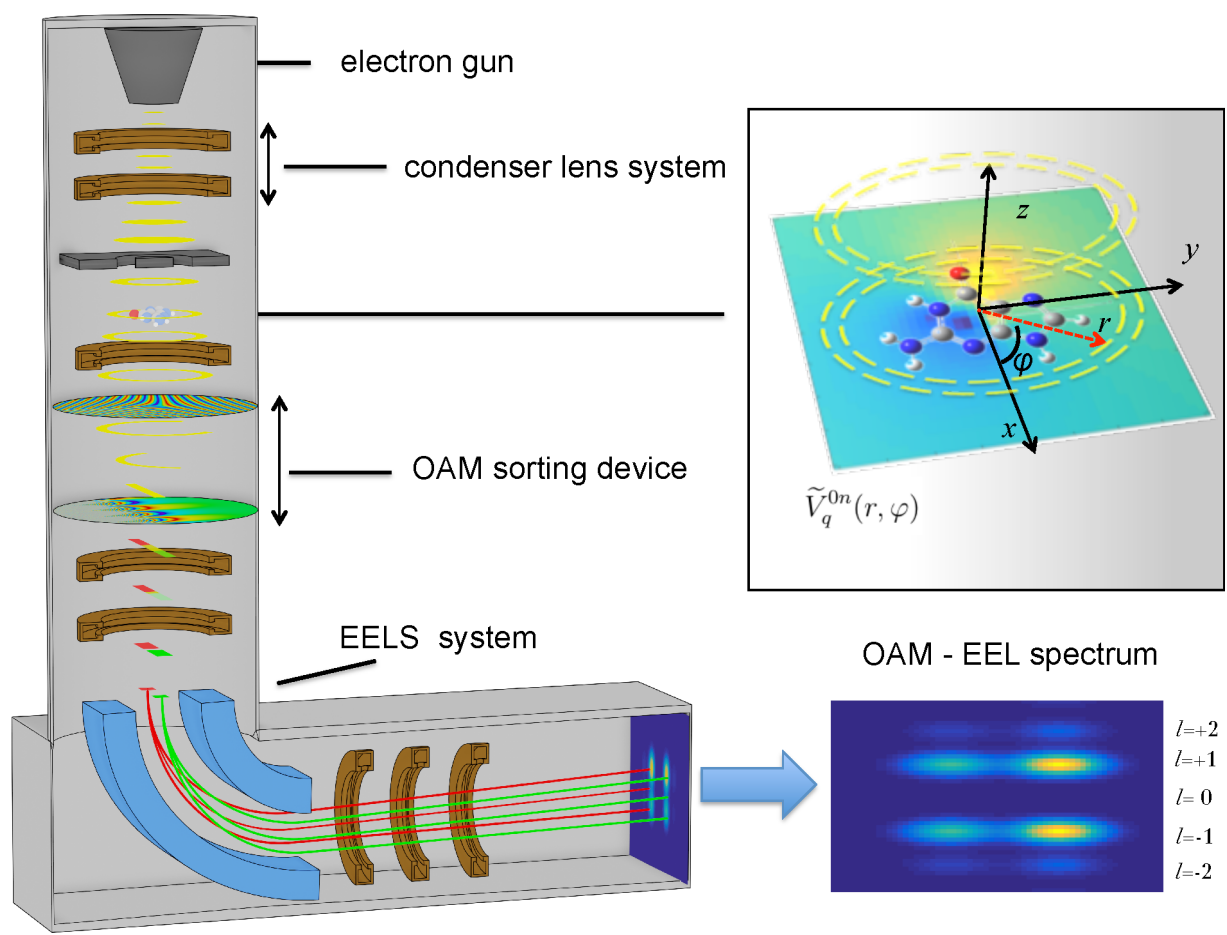

**Figure 1.** Scheme of an OAM-resolved EELS experiment to investigate a molecular system. The electron beam source system produces an annular shaped electron beam. The Coulombic interaction of the annular electron beam with the induced molecular transition potentials (inset, right panel) gives rise to the scattered electrons, processed in the EELS system, after the passage through an OAM sorting device. This last consist of two electrostatic phase elements in the electron column as detailed in refs 30, 39, and 40.

Our goal here is indeed to explore the possibility of conceiving new electron energy loss experiments for molecular and supramolecular systems. To obtain a new electronic excitation fingerprint, we propose to probe the azimuthal symmetry of the molecular transitions, based on the analysis of the different orbital angular momentum (OAM) components of the scattered electrons in TEM and STEM.[25,26] Indeed, free electrons can carry a quantized OAM value upon free-space propagation: these "*electron vortices*" are characterized by a spiraling wavefront with a screw dislocation along the propagation axis.[27,28] As a matter of fact, even if the measure of the OAM spectrum of a light beam was demonstrated experimentally 10 years ago,[29] only recently has the electronic analogue been made possible by devices based on electrostatic phase elements for measuring and spatially dispersing the different electrons' OAM components.[30,31] Our work is inspired by what has been proposed in the field of metallic nanostructures,[32] here extended to treat molecular and supramolecular systems. In the following, we describe how to modify the configuration of a TEM-EELS apparatus and how to encode a quantum chemistry treatment of the molecular systems and its interaction with the structured wave of the swift electron to obtain OAM resolved EELS spectra, then simulations of the expected experimental results will be presented in a number of paradigmatic cases considering also the effects of the finite resolutions in both energy and OAM due to a nonideal setup.

## ■ METHODS

**Beam Setup and Electron Optics Configuration.** A problem one can face in an EELS experiment performed on molecular systems is the avoidance of direct interaction of such highly energetic electrons with the specimen, which may substantially alter and destroy the structure of interest during observation. The use of aloof beam electron energy-loss spectroscopy as a nondestructive nanoscale surface characterization tool is one of the most powerful recent advances in this technique.[24,33,34] For instance, an aloof configuration of the beam, positioned tens of nanometers away from the sample, has been recently used for the detection of electronic and vibrational peaks in guanine crystals extracted from the scales of the Japanese koi fish:[24] by controlling the distance of an external narrow electron probe from the edge of the specimen, the authors selectively probe vibrational modes without exceeding the energy thresholds that potentially lead to radiation damage. Here, we propose that the control over

beam−sample interaction can be performed by an annular electron beam.[35−38]

One can imagine the experimental setup as depicted in Figure 1: an electron gun and a phase hologram in the condenser system[37,38] produce an annular shaped electron beam that interacts with the molecular specimen without hitting it.

The inelastically scattered electrons are sorted as a function of the different orbital angular momentum components using a set of two electrostatic phase elements in the electron column.[30−32,39,40] Interested readers can find a detailed description of this type of device in refs 39 and 40. Finally, the separated OAM components are processed by the EEL spectrometer system that produces the diffraction image observed on a fluorescent screen, giving rise to a double disperse spectrum as a function of the energies and angular momenta that is determined by the azimuthal symmetry of the molecular transition density probed (inset of Figure 1), as detailed below. Concerning the signal to noise ratio, one should note that, as in cryomicroscopy, this is linked to the maximum allowed dose.[24] The aloof configuration is more effective at reducing the damage for more delocalized processes, therefore at lower energy losses. While the present configuration would allow only a marginal gain of dose (order of 2) in the systems here reported, it will be much more effective for larger molecules and with lower energy losses: at losses of 1 eV or less and with an aloof scattering parameter of some nanometer distance, one can gain more than 1 order of magnitude in the allowed dose with a still decent signal.

**Transition Rate and Selection Rule.** Let us now describe how to properly model the electron−molecule interactions, to determine the final expected spectra. Atomic units are assumed to simplify the notation. A swift electron propagating in a homogeneous medium generates an electromagnetic field that can probe matter with a high spatial resolution. This field can be regarded as an evanescent source of radiation which permits exploring regions of momentum−energy space around the beam inducing electronic transitions in the target specimen,[11] from its ground state $|0\rangle$ of energy $E_0$ to generic excited states $|n\rangle$ of energy $E_n$.

Since the electrons are very energetic and the interaction can be considered to be generally small (at least compared to the kinetic energy of the beam electrons), the transition rate can be properly described within a first-order perturbation theory[11,20,41,42] (Fermi's golden rule-like). The initial (unperturbed) state $|\Psi_i\rangle$ of the electron plus molecule system can be written as $|\psi_i\rangle \otimes |0\rangle$; the state after the interaction is an electron-molecule entangled state that has acquired components $|\Psi_{fn}\rangle = |\psi_f\rangle \otimes |n\rangle$ on top of $|\Psi_i\rangle$. In other words, the incoming electron, described by its wave function $|\psi_i\rangle$ and energy $\varepsilon_i$, exchange energy, and momentum during the target−probe Coulomb interaction (that give rise to the coupling term) with the specimen, acquires components $|\psi_f\rangle$ of lower energy $\varepsilon_f$. Assuming that the molecule−swift electron interaction can be treated as purely electrostatic,[43] the probability for unit time $\Gamma^{EELS}(\omega)$ for the swift electron to lose an energy $\varepsilon_i - \varepsilon_f = \omega$ is given by

$$
\begin{aligned}
\Gamma^{EELS}(\omega) &= 2\pi \sum_f \sum_n |\langle \Psi_{fn}|\hat{H}_{int}|\Psi_i\rangle|^2 \, \delta(\omega - \omega_{0n}) \\
&= 2\pi \sum_f \sum_n \left| \int d\mathbf{r} \int d\mathbf{r}' \psi_f^*(\mathbf{r}) \frac{\langle n|\hat{\rho}(\mathbf{r}')|0\rangle}{|\mathbf{r} - \mathbf{r}'|} \psi_i(\mathbf{r}) \right|^2 \\
&\quad \times \delta(\omega - \omega_{0n}) \\
&= 2\pi \sum_f \sum_n |\langle \psi_f|\hat{H}'|\psi_i\rangle|^2 \, \delta(\omega - \omega_{0n})
\end{aligned}
\tag{1}
$$

where $\omega_{n0} = E_n - E_0$ are the excitation energies of the molecule and $\hat{\rho}(\mathbf{r}')$ is the electron density operator acting on the molecular electrons. The Coulomb coupling, $\hat{H}'(\mathbf{r})$, acts on the swift electron wave functions and is given by the interaction between an electron of the beam and the electrostatic potential due to the ground to excited state transition,[11,43] $V_{0n}(\mathbf{r})$:

$$
\hat{H}'(\mathbf{r}) = \int d^3\mathbf{r}' \frac{\langle n|\hat{\rho}(\mathbf{r}')|0\rangle}{|\mathbf{r} - \mathbf{r}'|} = \hat{V}_{0n}(\mathbf{r})
\tag{2}
$$

In our simulations, the molecular transition potential is calculated by adopting a linear response (LR) approach in the time-dependent density functional theory (TD-DFT) framework[44] (as detailed in the Supporting Information), but one can of course apply any appropriate electronic structure method that gives access to this quantity. Free-electron sources in electron microscopy generate unpolarized particles, which are described by the scalar wave function (in sharp contrast to optics) and are highly paraxial; i.e., the fields propagate along the direction of the free-electron motion $z$ and spread out only slowly in the transverse direction.[27] In these cases, the wavevectors $\mathbf{k} = (k_x, k_y, k_z)$ in the angular spectrum representation are almost parallel to the $z$ axis, and the transverse wavenumbers $(k_x, k_y)$ are small compared with $|\mathbf{k}| \approx k_z$. In the following, we apply this paraxial approximation to find a description of the individual electron wave functions. As consequence of the weak coupling and the ansatz of the wave functions, the initial and final states can be expressed as the product of a parallel and transverse component:

$$
\psi_i(\mathbf{r}) = \phi_i(\mathbf{r}_\perp) e^{ik_{iz}\cdot z}
\tag{3}
$$

$$
\psi_f(\mathbf{r}) = \phi_f(\mathbf{r}_\perp) e^{ik_{fz}\cdot z}
\tag{4}
$$

Here, $\mathbf{r}$ is the generic position vector of modulus $|\mathbf{r}|$. The annular component (in the plane perpendicular to the beam axis, $\mathbf{r}_\perp$) of the incident electron beam is set therefore[45]

$$
\phi_i(\mathbf{r}_\perp) = \frac{1}{N} e^{-\left(\frac{|\mathbf{r}| - r_0}{\Delta r}\right)^2}
\tag{5}
$$

with $N$ being the normalization constant, $r_0$ the annular internal radius, and $\Delta r$ the beam waist.

The final state is the direct product of the molecular excited state times the final free-electron one. Due to the aloof configuration, this last, eq 4, is assumed to be the product of a plane wave propagating in the parallel direction of the optical axis (and the beam, due to the paraxial approximation) with a transverse component expressed in cylindrical coordinates as a radial part and an azimuthal component:[28]

$$
\phi_f(\mathbf{r}_\perp) = J_{|l|}(k_{f\perp}r) e^{-il\varphi}
\tag{6}
$$

$J_{|l|}(k_{f\perp}r)$ is a Bessel function of the first kind of order $l$ (the angular quantum number), with transverse wavevector $k_{f\perp}$ (i.e., the projection on the $xy$ plane perpendicular to the TEM axis). The azimuthal component, $\exp(-il\varphi)$, describes the amount of OAM carried by the beam ($L_z = \hbar l$). The modes with $l \neq 0$ are also called *vortex* beams.[27]

The solutions of the free electron Schrödinger equation in cylindrical coordinates are a convenient basis due to the symmetry of the problem and to express the different OAM components of the scattered electron spectrum. In this way, the sum over the final electronic beam states appearing in eq 1 can be performed over an ensemble of such final states characterized by a fixed $l$ and transverse wave vector $k_{f\perp}$ in the range $[0, k_{max}]$. Here, $k_{max}$ is related to the collection angle of the detector ($\alpha$) by the de Broglie probe electron's wavelength ($\lambda$) as $k_{max} = \alpha\lambda$.

Concerning the molecular transition potentials associated with the electronic excitations, one can expect a sinusoidal (or cosinusoidal) azimuthal dependence, such as $\sin(m\varphi)$ or $\cos(m\varphi)$ (inset in Figure 1). Making use of the Fourier transform (FT) of transition potential $V_{0n}(\mathbf{r})$ along the longitudinal direction ($p = k_{fz} - k_{iz}$), we can therefore write down a transverse component of the transition potential:

$$
\begin{aligned}
\tilde{V}_p^{0n}(r, \varphi) &= \int V_{0n}(\mathbf{r})\, e^{-i(k_{fz}-k_{iz})\cdot z}\, dz \\
&= \int V_{0n}(\mathbf{r})\, e^{-ip\cdot z}\, dz \\
&\equiv \sum_{m=-\infty}^{+\infty} \tilde{V}_{p,m}^{0n}(r)\frac{(e^{-im\varphi} \pm e^{im\varphi})}{2}
\end{aligned}
\tag{7}
$$

Equation 7 makes use of the multipole expansion of the transition potential transverse component, expressed in cylindrical coordinates: $m = 0$ corresponds to the monopole, $m = \pm 1$ to the dipolar term, $m = \pm 2$ to the quadrupole, and so on (in general, the $2^{|m|}$ pole). The final energy loss rate per unit of angular momentum can be therefore conveniently re-expressed in cylindrical coordinates:

$$
\begin{aligned}
\Gamma_l^{EELS}(\omega) = \pi \int_0^{k_{max}} dk_f \\
\times \sum_n \left| \sum_{m=-\infty}^{+\infty} \int r\, dr \int_0^{2\pi} J_{|l|}(k_f r) e^{-il\varphi} \tilde{V}_{p,m}^{0n}(r) \right. \\
\left. (e^{-im\varphi} \pm e^{im\varphi}) e^{-\left(\frac{|\mathbf{r}|-r_0}{\Delta r}\right)^2} d\varphi \right|^2 \\
\times \delta(\omega - \omega_{0n})
\end{aligned}
\tag{8}
$$

It is now easy to derive the selection rule for this double resolved electron spectroscopy, i.e.:

$$
\int_0^{2\pi} e^{-i(l\pm m)\varphi}\, d\varphi \neq 0 \Rightarrow l = \mp m
\tag{9}
$$

Equations 9 and 10 show that by performing an OAM-EELS experiment one can simultaneously probe the energy and the different azimuthal components of the transition potential related to the different electronic excitations, due to the exchange of energy and momentum between individual electrons of the beam and the molecular system. The scattered electrons acquire all the different azimuthal components of the induced transition potentials: they are structured waves containing a multitude of vortices (one for each $l$ component acquired). The different OAM components can be sorted by the electrostatic optical elements.[30,31,39] As a consequence, also optically dark (i.e., dipole forbidden) transitions became detectable due to the signals of the higher angular momenta, i.e., those with $l \neq \pm 1$.

It is important to note that decentering the molecular system with respect to the beam axis can modify the selection rule in eq 10, in analogy to what was observed for atomic systems:[46] in the following, we therefore always consider cases where the optical axis is centered to the barycenter of the systems.

## ■ RESULTS AND DISCUSSION

In the following, we present some proof of concept simulations, performed at the DFT level and based on the theoretical description discussed in the previous section.

The geometrical structures have been optimized using the B3LYP[47] exchange-correlation (xc) functional and the cc-pVTZ basis set. We select guanine and guanine-based supramolecular assembly as prototype systems: we therefore make use of the same functional in all TD-DFT simulations to understanding the aggregation effects without including differences due to the electronic structure calculations.

In particular, the CAM-B3LYP[48] range-separated functional have been employed to correctly describe both local valence and charge-transfer excitations; indeed the order of the states can change by using different xc functionals, due to the charge transfer overstabilization problem of global hybrid functionals.[49] In addition, we also note that CAM-B3LYP has shown good performances in reproducing excitations of guanine compared to EOM-CCSD[50] and CASPT2[51] simulations and for the study of the excitonic properties of G-Quadruplexes with different tetrad stacking geometries.[50]

Concerning the basis set, we employed the cc-pVTZ basis also for the excited state simulations, with the exception of the G-quadruplex system for which the smaller 6-31G(d) basis set was used. Electronic structure simulations have been performed with the Gaussian16 version of the Gaussian suite of programs.[52] The transition potentials, stored on cubic grids, have been used to numerically integrate eq 9 by a Matlab script. We point out that the cubic grids here used are not the Gaussian 16 default but user-provided ones, with a very fine spacing of 0.1 Å. All the computational details are reported in the Supporting Information. To be useful as a spectroscopic tool, the supporting substrate should not perturb significantly the electronic structure of the molecular systems: therefore all the linear response calculations have been performed in vacuo. All simulations have been carried out keeping the collecting angle fixed to 200 mrad and the beam energy to 60 keV, whereas the loss probability has been determined in the angular momentum range of $[-3:3]$. In order to obtain doubled resolved spectra that simulate the limited instrumental resolution ($\Delta E = 0.1$ eV, $\Delta l = 0.5\hbar$), we have performed a convolution of OAM-EELS rates with the product of two Gaussian functions, following the procedure outlined in the Supporting Information of ref 32.

Let us start with testing a single molecule experiment, using guanine as a pedagogical case. Even if it will probably be difficult to experimentally perform a single molecule measure—at least in the first implementations of the experiment here proposed—we still find it instructive to start with this case and then increase the dimension of the specimen studied. The structure of guanine is shown in Figure 2, panel a, together with the transition potentials (integrated along $z$) of the first five singlet excitations (panel e). Assuming that the sample is illuminated by an annular beam with a radius of 7 au and a width of 3 a.u., we obtain the simulated OAM-EEL spectra for different values of OAM as reported in panels b, c, and d of Figure 2, using a Gaussian convolution with a broadening of 0.1 eV. As one can expect from the transition potential projections, all the transitions—except the dark singlet $S_4$—show a large dipolar component ($l = \pm 1$) that is obviously proportional to the corresponding optical oscillator strength (panel e, Figure 2). A small quadrupolar contribution is shown for the $S_5$ excitation and, to a lesser extent, also for those involving the $S_1$, $S_2$, and $S_3$ singlets (panel c, Figure 2). On the contrary, all the octupolar components are almost negligible (panels b and d, Figure 2). We note that the $S_4 \leftarrow S_0$ transition does not show any active component in the OAM range here considered ([−3:3]).

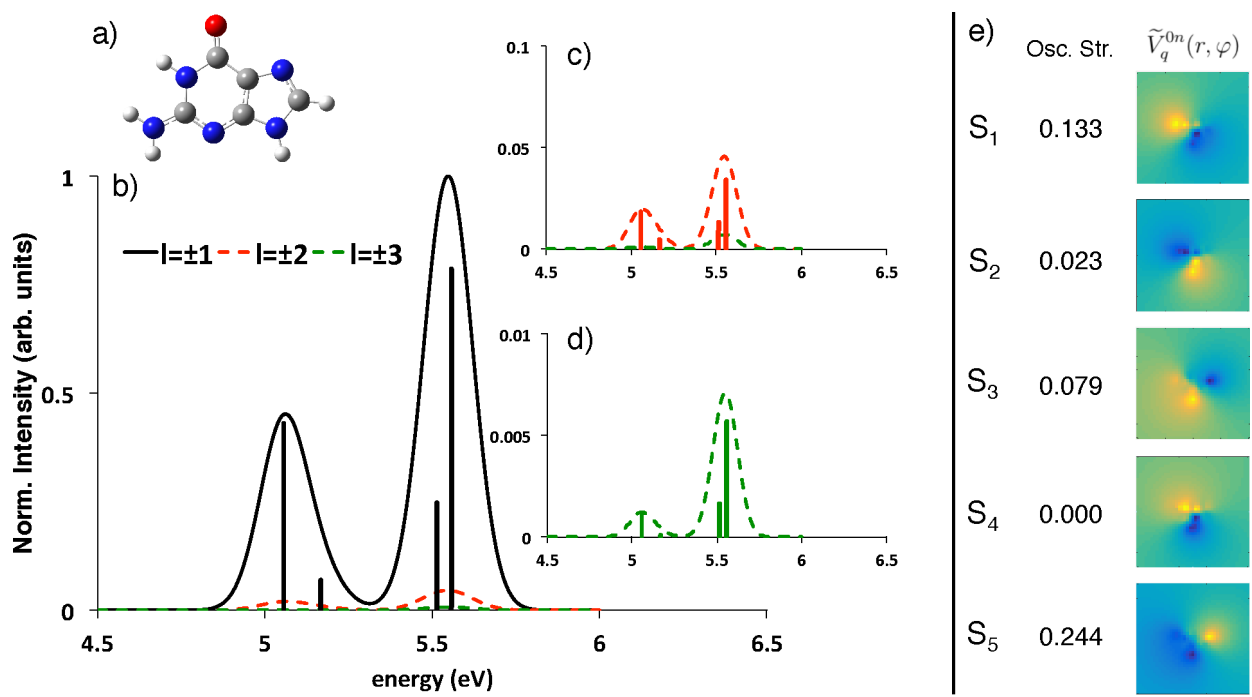

**Figure 2.** (a) Structure of optimized guanine. (b) Simulated OAM-resolved EEL spectra for different OAM values: black solid line $l = \pm 1$, red dashed line $l = \pm 2$, green dashed line $l = \pm 3$. Enlarged spectra for $l = \pm 2$ and $l = \pm 3$ are shown in figure insets c and d. For each spectrum, the stick components are reported. All spectra are normalized and have been convoluted with a Gaussian function simulating the limited instrumental resolution ($\Delta E = 0.1$ eV). (e) TD-DFT oscillator strength and transition potential projected along the $z$ direction for the first five singlet transitions.

Moving to more extended systems, we considered a tetramer of guanine bases, arranged in a planar configuration (Figure 3, panel a). The annular shaped electron beam used has a radius of 20 au and 3 au of beam waist. The geometry of such an arrangement has been obtained optimizing a monolayer of guanine on top of a gold slab with four layers of Au(111), as detailed by Rosa et al.;[53] then, the first eight singlet transitions were determined by CAM-B3LYP/6-311G(d) TD-DFT calculations. The geometry of each guanine monomer is slightly different from the optimized one used in the previous case; therefore we also performed simulations at the same level of theory on a monomer extracted from the tetramer. The states of the tetramer can indeed be described as excitonic combinations of the first two transitions of each monomer (at 5.06 and 5.36 eV respectively), as evident when inspecting panels b–e of Figure 3. The OAM-EEL spectrum for $l = \pm 1$ is indeed due to the convolution of two nearly degenerate states that are the result of the excitonic combinations of the first two transitions of each monomer. Each of them give rise indeed to two optically bright ($S_2$ and $S_3$; $S_6$ and $S_7$) and two optically dark ($S_1$ and $S_4$; $S_5$ and $S_8$) excitons, whose oscillator strengths are reported in panel f of Figure 3, together with the corresponding transition potential projections along $z$. A very important result comes out: two of the dark transitions ($S_1 \leftarrow S_0$ and $S_8 \leftarrow S_0$) can be indeed detected probing the different azimuthal symmetries of the corresponding potentials, in addition to the dipolar component, such as those corresponding to $l = \pm 2$. We note however that, as in the case of the single guanine, here the signals for the fourth and fifth singlets are too low to be recorded, limiting the sorting to the [−3:3] range. A very low but not negligible $l = \pm 4$ component is however observed for the $S_5 \leftarrow S_0$ transition (panel g and Table S13). Panel e of Figure 3 points out that, as expected, moving from the monomer to the tetramer causes an increase of the intensity and the approach of different peaks, induced by the aggregation. The enhancement is particularly large in the case of higher angular momenta: 2 and 1 order of magnitude for $l = \pm 2$ and $l = \pm 3$, respectively. This is very encouraging, because it shows that this technique can indeed be very effective to study extended supramolecular systems, such as those of biological interest. A double dispersed (energy and angular momenta) spectrum has finally been convoluted in panel g of Figure 3, to simulate what a possible experimental result should look like. Here, we assumed a limited instrumental resolution of $\Delta E = 0.1$–$0.3$ eV and $\Delta l = 0.5\hbar$.

As a final application, we focus on the study of chiral systems: as indeed pointed out in the field of nano-plasmonics,[41] electron OAM dichroism is expected for chiral systems in the presence of conventional electron beams. This means that difference in intensities between the loss functions $\Gamma_{+l}(\omega)$ and $\Gamma_{-l}(\omega)$ can be detected. The dichroism can be quantified, introducing a dichroic figure of merit:[41]

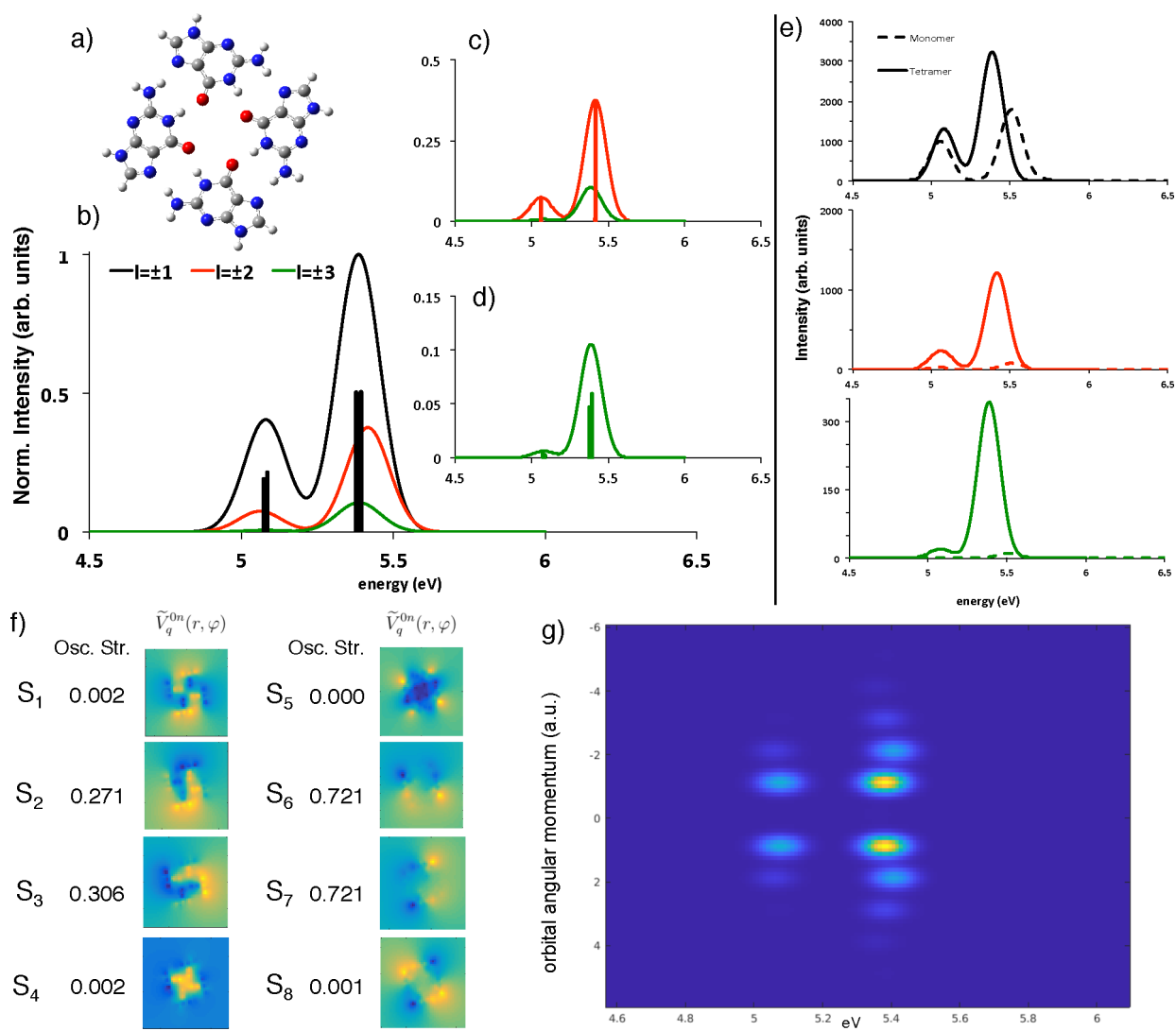

**Figure 3.** (a) Structure of a guanine tetramer, disposed in a planar configuration. (b) Simulated OAM-resolved EEL normalized spectra for different OAM values: black solid line $l = \pm 1$, red dashed line $l = \pm 2$, green dashed line $l = \pm 3$. Enlarged spectra for $l = \pm 2$ and $l = \pm 3$ are shown in figure insets c and d. For each spectrum, the stick components are reported. (e) Simulated OAM-resolved EEL spectra (not normalized) for different OAM values of the monomer (dashed lines) and tetramer (solid lines). The same color code of panels b–d has been applied. (f) TD-DFT oscillator strength and transition potential projected along the $z$ direction of the first five singlet transitions. (g) 2D representation of the OAM-EEL spectra simulating a realistic experiment. All spectra have been convoluted with a Gaussian function simulating the possible instrumental resolution ($\Delta E = 0.1$ eV, $\Delta l = 0.5\hbar$).

$$D_{|l|}(\omega) = \frac{\Gamma_{+l}(\omega) - \Gamma_{-l}(\omega)}{\Gamma_{+l}(\omega) + \Gamma_{-l}(\omega)} \times 100\% \qquad (10)$$

We note that this term is analogous (in percentage) to the dyssimetry factor of optical circular dichroism[54] (CD) experiments, apart from a factor of 2. As a first chiral paradigmatic case, we considered an amino acid, alanine, shown in its two enantiomeric forms, L and D, in Figure 4, panel a, whose double dispersed (energy and angular momenta) $\Gamma_l(\omega)$ spectrum is reported in panel b of the same figure. The annular shaped electron beam has the following dimensions: 7 au of radius and 3-a.u.-thick. From these data, the $D_{|l|}(\omega)$ can be obtained, and the convoluted spectra for $|l| = 1$ (Figure 4, panel c) point out how this technique is the electron beam analogue of an optical CD analysis: the two enantiomers give rise to two mirror spectra (this is of course true for all of the $|l|$ values, data not shown). Incidentally, it can be not easy to distinguish between two different enantiomers from the simple inspection of 2D spectra: we therefore suggest the use of the OAM-resolved EEL spectra and the calculation of the dichroic figure of merit in every experiment performed. Nevertheless, the very important result here is that one can probe the dichroism not only for the dipolar component, as in the optical CD, but can have access to all the different symmetries of the transitions potentials involved (Figure 4, panel d): absolute configurations and conformational analysis of molecular and supramolecular systems can be investigated taking into account also the contributions from dark transitions. Obviously, one has to note that in some cases even the optical CD can be

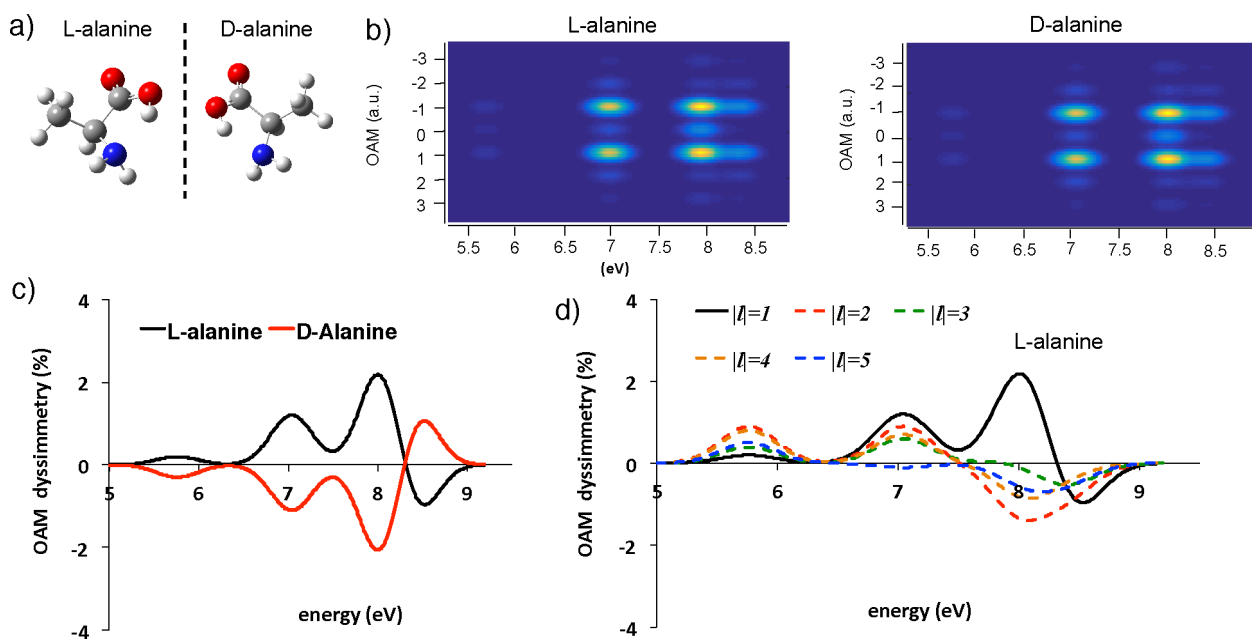

**Figure 4.** (a) Structure of the L and D enantiomers of alanine. (b) 2D representation of the OAM-EEL spectra simulating a realistic experiment. (c) Simulated spectra of OAM percentage of dichroism for $|l| = 1$ of both enantiomers. In black, the L-alanine; in red, the D-alanine. (d) Simulated spectra of OAM percentage of dichroism of L-alanine for different OAM values: black solid line $|l| = 1$, red dashed line $|l| = 2$, green dashed line $|l| = 3$, orange dashed line $|l| = 4$, blue dashed line $|l| = 5$. All spectra have been convoluted with Gaussian functions ($\Delta E = 0.33$ eV, $\Delta l = 0.5\hbar$).

somehow sensitive to "dark transitions," such as the magnetic-allowed, electric-forbidden ones.[55]

We therefore tested our protocol toward larger systems of biological interest, such as the G-quadruplex structures that originate in DNA and RNA guanine-rich sequences: the latter can indeed fold into tetra-helical structures stabilized by hydrogen bonds between guanine tetrads and electrostatic interactions with monovalent cations. Such arrangements, as a function of the specific sequence and folding conditions, can adopt various topologies classified in parallel and antiparallel depending on the relative direction of the four guanine strands. One can use optical CD spectroscopy to disentangle the two topologies, thanks to the particular fingerprints of the spectra: indeed, parallel G-quadruplexes are characterized by a positive couplet (two bands of opposite signs and similar amplitude; the sign of a couplet is defined by the longer-wavelength component) at 260 nm, whereas antiparallel G-quadruplexes have a positive band at 290 nm and a negative band at 260 nm.[56] Here, we selected one parallel (PDB code: 2MB2)[57] and one antiparallel (PDB code: 143D)[58] G-quadruplex, the core of which is formed by three guanine planes, as shown in panels a and b of Figure 5. The annular shaped electron beam used has a radius of 24 a.u. and 3 a.u. of beam waist. As shown in that figure, here only the cores of guanine chromophores have been taken into account to perform the simulations: the final geometries have been extracted from the NMR structures of PDB files, once refined by projecting the MP2/cc-pVDZ optimized geometry of the guanine base to the NMR structure, as detailed in ref 56. The two G-quadruplexes are not the enantiomers of the same compound; therefore the double dispersed OAM-EEL spectra, reported in panel c of Figure 3, are not expected to be the mirror image of the other (with respect to the $l = 0$ axis), contrary to what is observed in the L- and D-alanine cases. We simulated the spectra of $D_{l|l|}(\omega)$, convoluted with a Gaussian function (width = 0.1 eV, $|l| \in [0:3]$), in Figure 5, panel d. Even if for $|l| = 1$ no couplets are present, this is not the case of the higher OAM components: for $|l| = 2$, the parallel structure does not show appreciable dichroism (at least supposing a bandwidth of 0.1 eV), whereas the antiparallel configuration (143D) gives rise to a positive couplet around 5.02 eV. This last is mainly due to the second excitation in the positive part of the couplet and to the $S_6$, $S_8$, $S_{10}$, and $S_{15}$ in the negative one (see panel e of Figure 5). We note that the bands for $|l| = 3$ show an opposite trend for the two investigated structures: even if they cannot really be defined as a couplet because the two opposite peaks do not have a similar amplitude, however, in the case of the parallel structure, the positive band (at 5.08 eV) is very large and the negative one (at 5.29 eV) is very small; vice versa, the 143D system has an almost negligible positive peak at 4.93 eV and a large negative band at 5.23 eV (green lines, panel d of Figure 5).

## ■ CONCLUSIONS

In this contribution, we propose to extend the EELS technique to molecular and supramolecular systems making use of an electron beam configuration that avoids direct interaction with the specimen, and the sorting of scattered electrons as a function of the different electron orbital angular momenta. We have therefore demonstrated theoretically that, combining the evaluation of the energy and the OAM spectra of inelastically scattered electrons by (supra-)molecular systems, it is possible to obtain additional information about the symmetries and also the chirality of electronic excitations induced in these systems, by performing only a single measurement, i.e., without the need of modifying the features of the incoming electron wave. Indeed, thanks to the interaction with the molecular excitations, the scattered electrons acquire different components of helical phase fronts and spiralling currents, carrying a well-defined OAM per particle along the TEM axis. Finally, the numerical simulations performed clearly pointed out that this new technique, exploring the azimuthal symmetries of transitions, provides a unique molecular fingerprint that can be used to disentangle near degenerate states, to detect optically dark transitions, or to assign different topologies in extended systems.

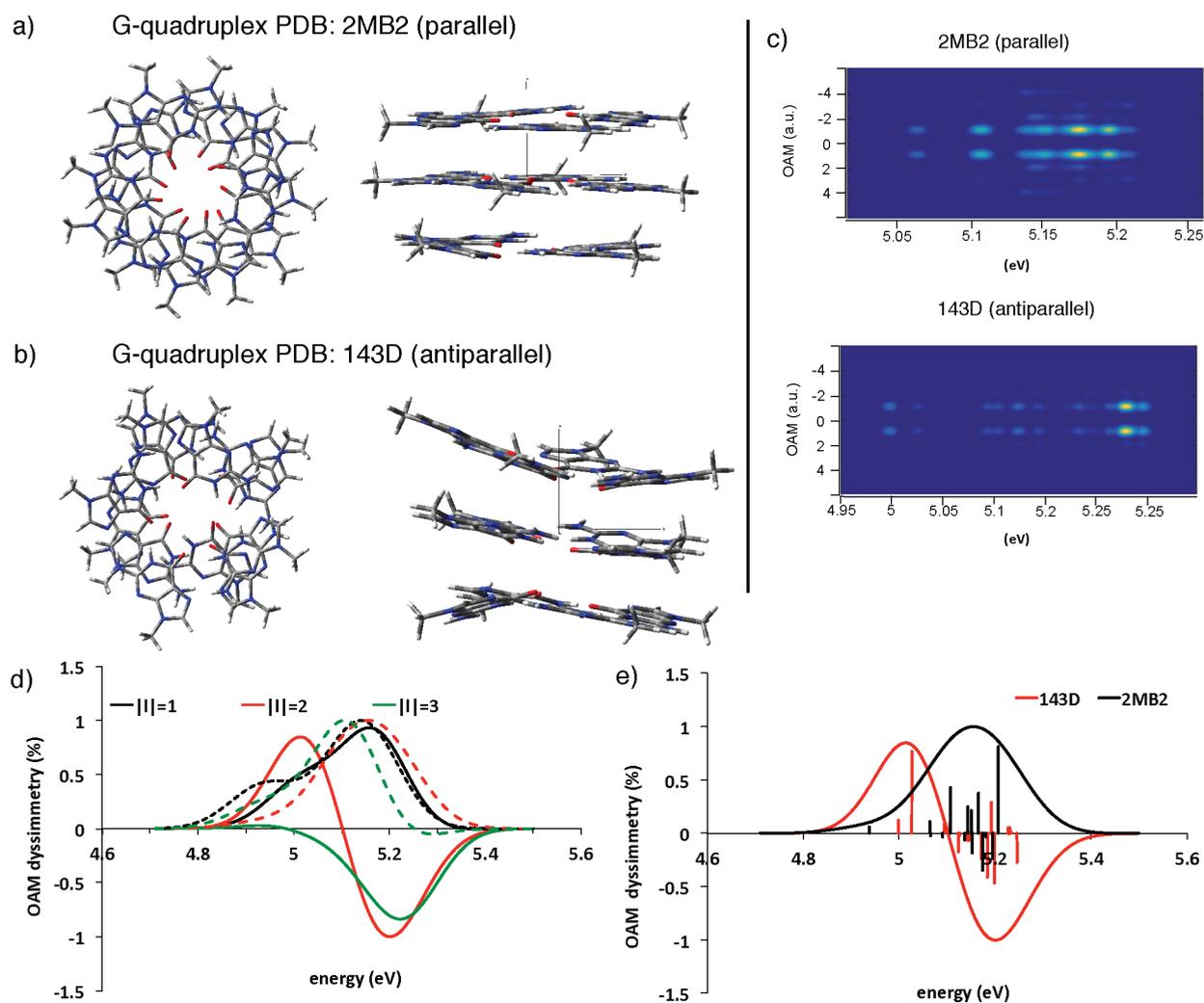

**Figure 5.** (a) Top and side view of the guanine core in G-quadruplex 2MB2. (b) Top and side view of the guanine core in G-quadruplex 143D. (c) 2D representation of the OAM-EEL spectra simulating a realistic experiment. Top: guanine core in G-quadruplex 2MB2. Bottom: guanine core in G-quadruplex 143D. (d) Simulated spectra (normalized) of OAM percentage of dichroism of guanine cores in 2MB2 G-quadruplex (dashed lines) and in the 143D one (solid lines), for different OAM values: black line $|l| = 1$, red line $|l| = 2$, green line $|l| = 3$. (e) Simulated spectra (normalized) of OAM percentage of dichroism for $|l| = 1$ of both G-quadruplexes. In black, the 2MB2 guanine core; in red, the 143D guanine core. Sticks at the different transitions are also reported. All spectra have been convoluted with Gaussian functions ($\Delta E = 0.1$ eV, $\Delta l = 0.5\hbar$).

As further extensions, we are now actively working to treat even more complex structures, such as pigment–protein complexes in light-harvesting (LH) systems.[59−61] A promising way is the integration of the theoretical framework here described with the excitonic description of the supramolecular system and the inclusion of the effect of environment.[62] Another active field is the study of the effects of plasmonic nanoantennae to enhance the intensity of the signal[63−65] for single molecule applications or the fine-tuning of the coherences in natural photosynthetic systems.[66,67] The present work can pave the way to many experimental applications in the field of biophysics and biochemistry: for instance, one could probe the role of optically dark charge transfer excitations in LH systems and their contribution in energy and electron transfer processes.[8,62]

## ■ ASSOCIATED CONTENT

### ⊕ Supporting Information

The Supporting Information is available free of charge at https://pubs.acs.org/doi/10.1021/acs.jctc.1c00045.

> Numerical integration procedure of the rate expression and the computational details of the simulated experiments (PDF)

## ■ AUTHOR INFORMATION

### Corresponding Authors

**Ciro A. Guido** − *Dipartimento di Scienze Chimiche, Università di Padova, 35131 Padova, Italy;* ⦿ orcid.org/0000-0003-1924-2862; Email: ciro.guido@unipd.it

**Enzo Rotunno** − *CNR-NANO, Institute of Nanoscience, Modena, Italy;* ⦿ orcid.org/0000-0003-1313-3884; Email: enzo.rotunno@nano.cnr.it

### Authors

**Matteo Zanfrognini** − *CNR-NANO, Institute of Nanoscience, Modena, Italy*

**Stefano Corni** − *Dipartimento di Scienze Chimiche, Università di Padova, 35131 Padova, Italy; CNR-NANO, Institute of Nanoscience, Modena, Italy;* ● orcid.org/0000-0001-6707-108X

**Vincenzo Grillo** − *CNR-NANO, Institute of Nanoscience, Modena, Italy*

Complete contact information is available at:
https://pubs.acs.org/10.1021/acs.jctc.1c00045

**Notes**
The authors declare no competing financial interest.

## ■ ACKNOWLEDGMENTS

C.A.G. thanks D.J. Fox and G. Scalmani of Gaussian Inc. for the useful hints and tips relating to the G16 software and M. Rosa (University of Padova) for providing the structure of the guanine tetramer slab. C.A.G. and S.C. acknowledge the support of the European Union's grant ERC-CoG-2015 No. 681285 "TAME-Plasmons" and the Italian Grant Agreement R164LZWZ4A MIUR-FARE Plasmo-Chem. E.R., M.Z., and V.G. acknowledge the support of the European Union's Horizon 2020 Research and Innovation Programme under Grant Agreement No. 766970 Q-SORT (H2020-FETOPEN-1-2016−2017).